\documentstyle[12pt]{article}
\newcommand{\zero}{{(0)}}
\newcommand{\be}{\begin{equation}}
\newcommand{\ee}{\end{equation}}
\newcommand{\ben}{\begin{eqnarray}\displaystyle}
\newcommand{\een}{\end{eqnarray}}

\newcommand{\wt}{\widetilde}
\newcommand{\wb}{\bar}
\newcommand{\refb}[1]{(\ref{#1})}

\newcommand{\LL}{{\cal L}}
\newcommand{\UU}{{\cal U}}
\newcommand{\MM}{{\cal M}}

\newcommand{\p}{\partial}

\newcommand{\ab}{{(a)}}
\newcommand{\bb}{{(b)}}
\newcommand{\bab}{{(\wb a)}}
\newcommand{\bbb}{{(\wb b)}}

\newcommand{\sectiono}[1]{\section{#1}\setcounter{equation}{0}}

\title{BLACK HOLE SOLUTIONS IN HETEROTIC STRING THEORY ON A TORUS}

\smallskip

\author{Ashoke Sen \\
Tata Institute of Fundamental Research \\
Homi Bhabha Road, Bombay 400005, INDIA \\
sen@theory.tifr.res.in, sen@tifrvax.bitnet
}

\begin{document}

\maketitle

\begin{abstract}

We construct the general electrically charged, rotating black hole solution
in the heterotic string theory compactified on a six dimensional torus and
study its classical properties.  This black hole is characterized by its
mass, angular momentum, and a 28 dimensional electric charge vector. We
recover the axion-dilaton black holes and Kaluza-Klein black holes for special
values of the charge vector.  For a generic black hole of this kind, the
28 dimensional magnetic dipole moment vector is not proportional to the
electric charge vector, and we need two different gyromagnetic ratios for
specifying the relation between these two vectors. We also give an algorithm
for constructing a 58 parameter rotating dyonic black hole solution in this
theory, characterized by its mass, angular momentum, a 28 dimensional electric
charge vector and a 28 dimensional magnetic charge vector. This is the most
general asymptotically flat black hole solution in this theory consistent
with the no-hair theorem.

\end{abstract}

\vfill

\vbox{\hbox{TIFR-TH-94-47}\hbox{hep-th/9411187}
\hbox{November, 1994}} \hfill ~

\eject

\sectiono{Introduction and Summary} \label{sone}

There has been a lot of activity in recent years towards construction of
black hole solutions in string theory\cite{REVIEWS}.
Since string theory is expected to
provide a finite quantum theory of gravity, we hope that within the
context of string theory we might be able to address various vexing
questions in black hole physics related to the black hole evaporation and
the consequent information loss puzzle. One suggestion in this
direction has been made in refs.\cite{SUSSKIND,RUSSO}.
It has also been proposed
there that massive elementary string states themselves should be identified
with black holes in string theory. Similar suggestions have also been made
in refs.\cite{HOOFT,DUFF,HULL} but for different reasons.

In order to study the physics of black holes in string theory,
we need to first construct the black hole solutions in string
theory, and then study their properties. In particular, study of the
relationship between black holes and elementary string states requires
us to construct the most general electrically charged black hole
solution in the theory.
In this paper we undertake the task of constructing the
most general black hole solution in one particular four dimensional
string theory, $-$ the heterotic string
theory compactified on a six dimensional torus. There are various reasons
for choosing this particular theory.
One of them is  that due to the existence of an N=4
supersymmetry in this theory, there are various non-renormalization
theorems which may make the quantum theory in this case more tractable
compared to the other compactification schemes. Second, this is perhaps
the best understood four dimensional string theory\cite{NARAIN}. And
finally, this theory is expected to possess a strong-weak coupling
duality symmetry\cite{SREV}
which may make the study of the non-perturbative physics more
feasible.

The theory under consideration has 28 U(1) gauge fields, and thus one
would expect that a generic electrically
charged black hole should be labelled by a 28 dimensional charge vector.
In section \ref{stwo} we use the technique of O(d,d+p)
transformations\cite{ODD1,ODD2,HASSAN,REVIEWS} to explicitly construct
a black hole solution characterized by an arbitrary 28 dimensional charge
vector, as well as mass and angular momentum, and
study its various properties.\footnote{Although the general procedure
for constructing such a solution was outlined in ref.\cite{SLTZ}, in the
present paper we find a simple form of the O(7,23) transformations on the
four dimensional fields that allows us to construct these solutions
explicitly.}
The relevant transformations in the present case generate the
group O(7,23).
The final solution is given in eqs.\refb{e26} - \refb{e34}.
For various special values of the parameters we can identify the solution
to various known solutions {\it e.g.} the rotating charged dilaton black
hole of ref.\cite{ROTAT} and the rotating Kaluza-Klein black hole of
refs.\cite{FROLOV,HORHOR}. One novel feature of a generic black hole
solution is that the 28 dimensional vector representing its magnetic
dipole moment is not, in general, parallel to the 28 dimensional
vector representing its electric charge, and we need two gyromagnetic
ratios for specifying the relation between these two vectors.
Also, a generic black hole
solution has all the 132 moduli fields, as well as the dilaton-axion
field non-zero.
We discuss the extremal limits, both
for non-zero angular momentum and zero angular momentum.
For non-rotating black holes, some of the extremal black holes saturate
the Bogomol'nyi bound and hence have unbroken supersymmetry. But
black holes carrying non-zero angular momentum never saturate Bogomol'nyi
bound, and hence have no unbroken supersymmetry.

If we allow the solution to carry magnetic charges also, then it
should be characterized by 58 parameters, the mass, the angular momentum,
28 electric charges, and 28 magnetic charges. In section \ref{sthree} we
give an algorithm to construct this 58 parameter rotating dyonic black
hole solution. This is done by using a combination of the S-duality
transformations\cite{STW,SLTZ} belonging to the group SL(2,R) and
the O(7,23) transformations\cite{BAKAS,KECH,BAKAS2,MYERS2,HULL}.
These two transformations together generate the
group O(8,24)\cite{MARSCH,THREED}.
This algorithm in fact naturally produces a 59 parameter
solution with singular metric,
belonging to the Taub-NUT family. We recover the non-singular
black hole solutions by imposing one restriction on these 59 parameters,
which sets the NUT charge to zero.

We conclude the paper in section \ref{sfour} by pointing out some
amusing coincidences between the properties of black holes and
those of elementary string states in the extremal limit.

\sectiono{Rotating Electrically Charged Black Holes} \label{stwo}

The massless fields in heterotic string theory compactified on a six
dimensional torus consist of the metric $G_{\mu\nu}$, the anti-symmetric
tensor field $B_{\mu\nu}$, twenty eight U(1) gauge fields $A_\mu^\ab$
($1\le a\le 28$), the scalar dilaton field $\Phi$, and a $28\times 28$
matrix valued scalar field $M$ satisfying,
\be \label{e5}
M L M^T = L, \quad \quad M^T=M.
\ee
Here $L$ is a 28$\times$28 symmetric matrix with 22 eigenvalues $-1$ and
6 eigenvalues $+1$. For definiteness we shall take $L$ to be,
\be \label{e4}
L=\pmatrix{-I_{22} & \cr & I_6\cr},
\ee
where $I_n$ denotes an $n\times n$ identity matrix. The action describing
the effective field theory of these massless
bosonic fields is given by\cite{MAHSCH},
\ben \label{e1}
S &=& C\int d^4 x \sqrt{-\det G} \, e^{-\Phi} \, \Big[ R_G + G^{\mu\nu}
\p_\mu \Phi \p_\nu\Phi +{1\over 8} G^{\mu\nu} Tr(\p_\mu M L\p_\nu ML)
\nonumber \\
&& -{1\over 12} G^{\mu\mu'} G^{\nu\nu'} G^{\rho\rho'} H_{\mu\nu\rho}
H_{\mu'\nu'\rho'} - G^{\mu\mu'} G^{\nu\nu'} F^\ab_{\mu\nu} \, (LML)_{ab}
\, F^\bb_{\mu'\nu'} \Big] \, ,
\een
where,
\be \label{e2}
F^\ab_{\mu\nu} = \p_\mu A^\ab_\nu - \p_\nu A^\ab_\mu \, ,
\ee
\be \label{e3}
H_{\mu\nu\rho} = \p_\mu B_{\nu\rho} + 2 A_\mu^\ab L_{ab} F^\bb_{\nu\rho}
+\hbox{cyclic permutations of $\mu$, $\nu$, $\rho$}\, ,
\ee
and $R_G$ denotes the scalar curvature associated with the metric
$G_{\mu\nu}$.
$C$ is an arbitrary constant which does not affect the equations of
motion and can be absorbed into the dilaton field $\Phi$.
This action is invariant under an O(6,22) transformation,
\be \label{e6}
M\to \Omega M \Omega^T, \quad \quad A_\mu^\ab \to \Omega_{ab}
A_\mu^\bb, \quad \quad \Phi\to \Phi, \quad \quad G_{\mu\nu}\to G_{\mu\nu},
\quad \quad B_{\mu\nu} \to B_{\mu\nu},
\ee
where $\Omega$ is a $28\times 28$ matrix satisfying,
\be \label{e6a}
\Omega L \Omega^T = L\, .
\ee
This is a symmetry of the full string theory if we also rotate the 28
dimensional lattice $\Lambda$ of electric charges by $\Omega$\cite{NARAIN}.
On the other hand, for fixed $\Lambda$, only a discrete subgroup O(6,22;Z)
of O(6,22) is a symmetry of the full string theory.

Let us now consider backgrounds that are independent of the time coordinate
$t$. In this case the action is expected to have an
O(7,23) symmetry\cite{HASSAN,REVIEWS}.
To see how this appears, let us define new variables as
follows:
\ben \label{e7}
\wb A^\ab_i &=& A^\ab_i - (G_{tt})^{-1} G_{ti} A^\ab_t\, , \quad
\quad 1\le a\le 28\, , \quad 1\le i\le 3\, , \nonumber \\
\wb A^{(29)}_i &=& {1\over 2} (G_{tt})^{-1} G_{ti} \, , \nonumber \\
\wb A^{(30)}_i &=& {1\over 2} B_{ti} + A^\ab_t L_{ab} \wb A^\bb_i\, ,
\een
\ben \label{e8}
\wb M &=& \pmatrix{M + 4(G_{tt})^{-1} A_t A_t^T & -2 (G_{tt})^{-1} A_t &
2ML A_t + 4(G_{tt})^{-1} A_t (A_t^T L A_t)\cr
&&\cr
- 2 (G_{tt})^{-1} A^T_t & (G_{tt})^{-1} & -2 (G_{tt})^{-1} A_t^TL A_t\cr
&&\cr
2 A_t^TLM + 4(G_{tt})^{-1} A_t^T (A_t^TLA_t) & -2(G_{tt})^{-1} A_t^T
L A_t & G_{tt} + 4 A_t^T LML A_t \cr
&& + 4(G_{tt})^{-1} (A^T_t L A_t)^2\cr}
\, ,  \nonumber \\  &&
\een
\be \label{e9}
\wb G_{ij} = G_{ij} - (G_{tt})^{-1} G_{ti} G_{tj}\, ,
\ee
\be \label{e10}
\wb B_{ij} = B_{ij} +(G_{tt})^{-1} (G_{ti}A^\ab_j - G_{tj} A^\ab_i)
L_{ab} A^\bb_t +{1\over 2} (G_{tt})^{-1} (B_{ti}G_{tj} - B_{tj} G_{ti})\, ,
\ee
\be \label{e11}
\wb \Phi = \Phi - {1\over 2} \ln (-G_{tt})\, ,
\ee
and,
\be \label{e12}
\wb L = \pmatrix{ L &0 &0 \cr 0&0& 1 \cr 0&1 &0 \cr}\, .
\ee
It can be verified that for time independent field configurations, the
action \refb{e1} can be rewritten as,
\ben \label{e13}
S &=& C\int dt \int d^3 x \sqrt{\det \wb G} e^{-\wb \Phi} \Big[ R_{\wb G}
+ \wb G^{ij}
\p_i \wb\Phi \p_j\wb \Phi +{1\over 8} \wb G^{ij} Tr(\p_i \wb M \, \wb L\p_j
\wb M \, \wb L)
\nonumber \\
&& -{1\over 12} \wb G^{ii'} \wb G^{jj'} \wb G^{kk'} \wb H_{ijk}
\wb H_{i'j'k'} - \wb G^{ii'} \wb G^{jj'} \wb F^\bab_{ij} (\wb L\wb M
\wb L)_{\wb a\wb b}\wb F^\bbb_{i'j'} \Big] \, ,
\een
where,
\be \label{e14}
\wb F^\bab_{ij} = \p_i \wb A^\bab_j - \p_j \wb A^\bab_i \, ,
\qquad 1\le \wb a \le 30,
\ee
and,
\be \label{e15}
\wb H_{ijk} = \p_i \wb B_{jk} + 2 \wb A_i^\bab \wb L_{\wb a\wb b} \wb
F^\bbb_{jk}
+\hbox{cyclic permutations of $i$, $j$, $k$}\, .
\ee
This action has an O(7,23) symmetry:
\be \label{e16}
\wb M\to \wb \Omega \wb M \wb \Omega^T, \quad \quad \wb A_i^\bab \to \wb
\Omega_{\wb a\wb b}
\wb A_i^\bbb, \quad \quad \wb\Phi\to \wb \Phi, \quad \quad \wb G_{ij}\to
\wb G_{ij},
\quad \quad \wb B_{ij} \to \wb B_{ij},
\ee
where $\wb\Omega$ is a $30\times 30$ matrix satisfying,
\be \label{e17}
\wb \Omega \wb L \wb \Omega^T = \wb L\, .
\ee
In order to get a convenient parametrization of $\wb\Omega$ it is easier
to work with the diagonal form of $\wb L$. The orthogonal
matrix $U$ that diagonalizes $\wb L$ is given by,
\be \label{e18}
U=\pmatrix{ I_{28} && \cr & {1\over \sqrt 2} & {1\over \sqrt 2} \cr
& {1\over \sqrt 2} & -{1\over \sqrt 2} \cr} \, .
\ee
We have,
\medskip
\be \label{e19}
U \wb L U^T \equiv \wb L_d = \pmatrix{ -I_{22} &&& \cr & I_6 && \cr && 1 &\cr
&&& -1\cr} \, .
\ee
Then $U \wb \Omega U^T$ preserves $\wb L_d$.

We can apply this O(7,23) transformation on a known time independent
classical solution to generate new classical solutions of the equations
of motion.
We shall restrict ourselves to solutions
characterized by fixed asymptotic  configuration of various fields,
representing asymptotically flat space time. For definiteness, we shall
look for solutions with the following asymptotic forms for various fields:
\medskip
\be \label{e20}
M_{as} = I_{28}, \quad \quad \Phi_{as} = 0, \quad \quad (A_\mu^\ab)_{as}
=0, \quad \quad
(G_{\mu\nu})_{as}=\eta_{\mu\nu}, \quad \quad (B_{\mu\nu})_{as} = 0\, .
\ee
This gives,
\medskip
\be \label{e21}
\wb M_{as} = \pmatrix{I_{28} && \cr & -1 & \cr && -1\cr}\, .
\ee
Given a solution with any other asymptotically constant
field configuration, we can bring it to the form \refb{e20}
by a combination of the O(6,22) transformation, the
transformation $\Phi\to \Phi+constant$, the general coordinate
transformation
involving linear change in the coordinates, and the freedom of shifting
$A_\mu^\ab$ and $B_{\mu\nu}$ by constants.
Thus we do not suffer from any loss of
generality by restricting to field configurations with asymptotic
behaviour given in eq.\refb{e20}.

As in ref.\cite{ROTAT}, we begin with the Kerr solution:
\ben \label{e22}
ds^2 &\equiv& G_{\mu\nu} dx^\mu dx^\nu \nonumber \\
&=& -{\rho^2 + a^2 \cos^2\theta - 2m\rho\over \rho^2 + a^2 \cos^2 \theta}
dt^2 +{\rho^2 + a^2 \cos^2 \theta \over \rho^2 + a^2 -2m \rho} d\rho^2
+ (\rho^2 + a^2 \cos^2\theta) d\theta^2 \nonumber \\
&& + {\sin^2\theta\over \rho^2 + a^2 \cos^2\theta} [(\rho^2+a^2) ( \rho^2
+ a^2 \cos^2\theta) + 2m\rho a^2\sin^2\theta] d\phi^2 \nonumber \\
&& -{4m\rho a \sin^2\theta \over \rho^2+a^2\cos^2\theta} dt d\phi
\nonumber \\
\Phi = 0, && \qquad B_{\mu\nu}=0, \qquad
A^\ab_\mu =0, \qquad M=I_{28}\, .
\een
Here $t$, $\rho$, $\theta$, $\phi$ denote the space-time coordinates.
This is guaranteed to be a solution of the equations of motion derived
from action \refb{e1} since when the $\Phi$, $B_{\mu\nu}$ and $A_\mu^\ab$
fields are set to zero, and $M$ is set to the identity matrix,
the equations of motion derived from the action
\refb{e1} become identical to Einstein's equation in matter free
space. Using eqs.\refb{e7}-\refb{e11} we get,
\ben \label{e22a}
&& \wb A^\bab_\phi = \delta_{\wb a, 29} \, {m\rho a \sin^2\theta
\over \rho^2 + a^2\cos^2\theta - 2m\rho}\, , \qquad
\wb A^\bab_\theta=0\, , \qquad \wb A^\bab_\rho=0\, , \nonumber \\
&&\wb M = \pmatrix{I_{28} && \cr & -f^{-1} & \cr &&
-f\cr}\, , \quad \quad f = {\rho^2 + a^2 \cos^2\theta - 2m\rho \over
\rho^2 + a^2 \cos^2\theta}\, , \nonumber \\
&&\wb G_{ij} dx^i dx^j = (\rho^2 + a^2\cos^2\theta) \Big[
{1 \over \rho^2 + a^2 -2m\rho} d\rho^2 + d\theta^2 + {\rho^2 + a^2 -2m\rho
\over \rho^2 + a^2 \cos^2\theta - 2m\rho} \sin^2\theta d\phi^2\Big]\, ,
\nonumber \\
&& \wb B_{ij} = 0, \qquad \wb \Phi = -{1\over 2} \ln f\, .
\een
We can now generate new solutions by
performing an O(7,23) transformation on this solution. Since, however,
we want to keep the asymptotic forms of various field configurations
fixed, we only use a subgroup of the O(7,23) transformations which
preserve \refb{e21}. This leaves us with an O(6,1)$\times$O(22,1)
subgroup of the full O(7,23) group.
If we describe the transformation by the matrix
$U\wb\Omega U^T$ instead of $\wb\Omega$, then
the O(6,1) subgroup acts on the
23rd - 28th, and the 30th index of the matrix $U \wb M U^T$, whereas the
O(22,1) subgroup acts on the 1st - 22nd, and the 29th index of the matrix
$U \wb M U^T$. Not every element of this O(6,1)$\times$O(22,1) subgroup
generates a new solution however. It is clear from eq.\refb{e22a} that an
O(22)$\times$O(6) subgroup, which acts on the 1st - 22nd and the 23rd
- 28th indices respectively, leaves the solution invariant.
Thus the transformations
which generate inequivalent solutions, preserving the asymptotic field
configuration, can be parametrized by the coset
\be \label{e22b}
\big(O(6,1)\times O(22,1)\big)/\big(O(6)\times O(22)\big)\, .
\ee
This is a 28 dimensional space.
We now need to find a convenient representative of a generic element
of this coset in the O(6,1)$\times$O(22,1) group. This is done as
follows. We take $\wb\Omega$ to be of the form:
\be \label{e22c}
\wb \Omega = \wb\Omega_2 \wb \Omega_1\, ,
\ee
where,
\be \label{e23}
U\wb \Omega_1 U^T = \pmatrix{ I_{21} & 0 & 0 & 0 & 0 & 0 \cr
0 & \cosh\alpha & 0 & 0 & \sinh\alpha & 0 \cr 0 & 0 & I_5 & 0 & 0 & 0 \cr
0 & 0 & 0 & \cosh\beta & 0 & \sinh\beta \cr
0 & \sinh\alpha & 0 & 0 & \cosh\alpha & 0 \cr
0 & 0 & 0 & \sinh\beta & 0 & \cosh\beta \cr}\, ,
\ee
and,
\be \label{e25}
\wb \Omega_2 = \pmatrix{R_{22}(\vec n) && \cr & R_6(\vec p)
& \cr && I_2\cr}\, , \ee
where $R_N(\vec k)$ denotes any $N$-dimensional rotation matrix that
rotates an $N$-dimensional column
vector with only $N$-th component non-zero and
equal to 1 to
an arbitrary $N$-dimensional unit vector $\vec k$. Here $\vec n$ and $\vec p$
are arbitrary $22$ and $6$ dimensional unit vectors respectively. Thus $\wb
\Omega$ given by eqs.\refb{e22c}-\refb{e25} is parametrized by 28
parameters $\alpha$, $\beta$, $\vec n$ and $\vec p$.

It is now a straightforward algebraic exercise to apply the transformation
generated by $\wb\Omega$ to the field configuration given in
eqs.\refb{e22}, \refb{e22a} and
extract the expressions for various fields in the transformed solution. The
result is,
\ben \label{e26}
ds^2 &\equiv& G_{\mu\nu} dx^\mu dx^\nu \nonumber \\
&=&  (\rho^2+a^2\cos^2\theta)\Big\{ - \Delta^{-1}
(\rho^2+a^2\cos^2\theta -2m\rho) dt^2 + (\rho^2+a^2-2m\rho)^{-1} d\rho^2
+d\theta^2 \nonumber \\
&& +\Delta^{-1} \sin^2\theta [\Delta + a^2\sin^2\theta (\rho^2 + a^2\cos^2
\theta + 2m\rho \cosh\alpha\cosh\beta)] \, d\phi^2
\nonumber \\ &&
- 2\Delta^{-1} m\rho a \sin^2\theta (\cosh\alpha + \cosh\beta) dt
d\phi\Big\}\, ,
\een
where,
\be \label{e27}
\Delta = (\rho^2 + a^2\cos^2\theta)^2 + 2m \rho (\rho^2 + a^2 \cos^2\theta)
(\cosh\alpha \cosh\beta -1) + m^2\rho^2 (\cosh\alpha -\cosh\beta)^2\, ,
\ee
\be \label{e28}
\Phi={1\over 2} \ln {(\rho^2+a^2\cos^2\theta)^2\over \Delta}\, ,
\ee
\ben \label{e29}
A^\ab_t &=& -{n^\ab\over \sqrt 2} \Delta^{-1} m\rho\sinh\alpha \{ (\rho^2+
a^2\cos^2\theta)\cosh\beta + m\rho (\cosh\alpha - \cosh\beta)\} \quad
\hbox{for} \, \, 1\le a\le 22\, , \nonumber \\
&=& -{p^{(a-22)}\over \sqrt 2} \Delta^{-1} m\rho\sinh\beta \{ (\rho^2+
a^2\cos^2\theta)\cosh\alpha + m\rho (\cosh\beta - \cosh\alpha)\} \quad
\hbox{for} \, \, a\ge 23\, , \nonumber \\
\een
\ben \label{e30}
A^\ab_\phi &=& {n^\ab\over \sqrt 2} \Delta^{-1} m \rho a \sinh\alpha
\sin^2\theta \{ \rho^2 + a^2\cos^2\theta + m\rho\cosh\beta (\cosh
\alpha -\cosh\beta)\}  \quad  1\le a\le 22\, , \nonumber \\
&=& {p^{(a-22)}\over \sqrt 2} \Delta^{-1} m \rho a \sinh\beta
\sin^2\theta \{ \rho^2 + a^2\cos^2\theta + m\rho\cosh\alpha (\cosh
\beta -\cosh\alpha)\} \quad  a\ge 23\, , \nonumber \\
\een
\be \label{e31}
B_{t\phi} = \Delta^{-1} m\rho a \sin^2\theta (\cosh\alpha -\cosh\beta)
\{ \rho^2 + a^2\cos^2\theta + m\rho(\cosh\alpha \cosh\beta -1)\} \,
\nonumber \\
\ee
\be \label{e32}
M = I_{28} +\pmatrix{ P nn^T & Q n p^T \cr Q p n^T & P pp^T\cr}\, ,
\ee
where,
\ben \label{e33}
P &=& 2\Delta^{-1} m^2\rho^2 \sinh^2\alpha \sinh^2\beta \, ,\nonumber \\
Q &=& - 2 \Delta^{-1} m\rho \sinh\alpha \sinh\beta \{ \rho^2 + a^2
\cos^2\theta + m\rho (\cosh\alpha \cosh\beta -1) \} \, .
\een
Note that the solution is characterized by non-trivial
dilaton as well as other scalar moduli fields $M$.
{}From eqs.\refb{e26} and \refb{e28} we can also find an expression for
the canonical Einstein metric $g_{\mu\nu}\equiv e^{-\Phi} G_{\mu\nu}$:
\ben \label{e34}
ds_E^2 &\equiv& g_{\mu\nu} dx^\mu dx^\nu = e^{-\Phi} ds^2 \nonumber \\
&=& \Delta^{1\over 2} \Big\{ - \Delta^{-1}
(\rho^2+a^2\cos^2\theta -2m\rho) dt^2 + (\rho^2+a^2-2m\rho)^{-1} d\rho^2
+d\theta^2 \nonumber \\
&& +\Delta^{-1} \sin^2\theta [\Delta + a^2\sin^2\theta (\rho^2 + a^2\cos^2
\theta + 2m\rho \cosh\alpha\cosh\beta)] \, d\phi^2
\nonumber \\  &&
- 2\Delta^{-1} m\rho a \sin^2\theta (\cosh\alpha + \cosh\beta) dt
d\phi\Big\}\, .
\een

One can easily verify that the new solution given in
eqs.\refb{e26}-\refb{e34} describes a black hole with mass $M$, angular
momentum $J$, electric charge $Q^\ab$ and magnetic dipole moment
$\mu^\ab$ given by,
\be \label{e35}
M = {1\over 2} m ( 1 + \cosh\alpha \cosh \beta)\, ,
\ee
\be \label{e36}
J = {1\over 2} ma (\cosh\alpha + \cosh \beta) \, ,
\ee
\ben \label{e37}
Q^\ab &=& {m\over \sqrt 2} \sinh\alpha \cosh\beta \, n^\ab \qquad \hbox{for}
\quad 1\le a \le 22\, \nonumber \\
&=& {m\over \sqrt 2} \sinh\beta \cosh\alpha \, p^{(a-22)} \qquad \hbox{for}
\quad 23\le a\le 28\, ,
\een
\ben \label{e38}
\mu^\ab &=& {1\over \sqrt 2} ma \sinh\alpha \, n^\ab \qquad \hbox{for}
\quad 1\le a \le 22\, \nonumber \\
&=& {1\over \sqrt 2} ma \sinh\beta \, p^{(a-22)} \qquad \hbox{for}
\quad 23\le a\le 28\, .
\een
{}From eqs.\refb{e37} and \refb{e38} we see that for generic values of
the parameters $\alpha$, $\beta$, $\vec n$ and $\vec p$, the 28-
dimensional vectors
$\vec \mu$ and $\vec Q$ are not parallel to each other.
The special cases for
which $\vec \mu$ and $\vec Q$ {\it are} parallel are i) $\beta=0$, ii)
$\alpha=0$, and iii) $\alpha=\beta$. The black hole solution in case (i)
corresponds to the rotating charged black hole solution discussed in
ref.\cite{ROTAT}, whereas the case $\alpha=\beta$ corresponds to the
Kaluza-Klein black hole discussed in refs.\cite{FROLOV,HORHOR}. To
check that the results \refb{e37} and \refb{e38} are consistent with the
ones derived in refs.\cite{ROTAT,HORHOR}, note that the gyromagnetic
ratio, defined as $2\mu M/QJ$, is equal to 2 in case (i), and is equal to
$(1+\hbox{sech}^2\alpha)$ in case (iii), which varies
between 1 and 2 as $\alpha$ varies between $\infty$ and 0.

Even though $\vec \mu$ and $\vec Q$ are not parallel in general, and hence
we cannot define an overall gyromagnetic ratio, we can define two separate
gyromagnetic ratios in the left and the right hand sector as follows.
Let us define
\be \label{e51}
Q^\ab_{L\atop R} = {1\over 2} (I_{28}\mp L)_{ab} Q^\bb\, , \quad \quad
\mu^\ab_{L\atop R} = {1\over 2} (I_{28}\mp L)_{ab} \mu^\bb\, .
\ee
Then $\vec Q_L$ ($\vec \mu_L$) has its last six components zero, whereas
$\vec Q_R$ ($\vec \mu_R$) has its
first 22 components zero. In other words, $\vec Q_L$ ($\vec \mu_L$)
and $\vec Q_R$ ($\vec \mu_R$) can
be regarded as 22 and 6 dimensional vectors respectively.
Then from eqs.\refb{e35} - \refb{e38} we get,
\be \label{emu1}
\vec \mu_L = {1\over 2} g_L {J\over M} \vec Q_L\, , \qquad
\vec \mu_R = {1\over 2} g_R {J\over M} \vec Q_R\, ,
\ee
where,
\be \label{emu2}
g_L = 2 \, {1+\cosh\alpha\cosh\beta\over \cosh\alpha+\cosh\beta}\,  {1
\over \cosh\beta} \, , \quad \quad
g_R = 2 \, {1+\cosh\alpha\cosh\beta\over \cosh\alpha+\cosh\beta}\, {1
\over \cosh\alpha} \, .
\ee

Various geometrical properties of the solution can be studied using the
canonical metric given in eq.\refb{e34}. There are two horizons
corresponding to the surfaces,
\be \label{e39}
\rho^2 -2m\rho + a^2 =0\, ,
\ee
which gives the location of the horizons at,
\be \label{e40}
\rho = m \pm \sqrt{m^2 - a^2} \equiv \rho_H^\pm \, .
\ee
Note that the horizons disappear for $a>m$, leaving behind naked
singularity. The limit $a\to m$ is known as the extremal limit.

The area of the outer event horizon, which is proportional to the
Bekenstein entropy of the black hole, is given by,
\be \label{e41}
A = \int d\theta d\phi \sqrt{g_{\theta\theta} g_{\phi\phi}}\, |_{\rho
=\rho_H^+} = 4\pi m (\cosh\alpha +\cosh\beta) \, (m+\sqrt{m^2 - a^2})\, .
\ee
The surface gravity of the black hole, calculated at $\theta=0$, is
given by,
\be \label{e42}
\kappa = \lim_{\rho\to \rho_H^+} \sqrt{g^{\rho\rho}} \p_\rho
\sqrt{-g_{tt}}\, |_{\theta=0} = { \sqrt{m^2 - a^2} \over
m(\cosh\alpha + \cosh\beta) (m+\sqrt{m^2 - a^2}) }\, .
\ee
$\kappa/2\pi$ can be interpreted as the Hawking temperature of the black
hole.  Finally, the angular velocity $\Omega$ at the horizon is found by
demanding that the vector ${\p\over \p t} + \Omega {\p\over \p \phi}$
is null at the horizon. This gives,
\be \label{e43}
g_{tt} + 2 g_{t\phi} \Omega + g_{\phi\phi} \Omega^2 = 0\, .
\ee
The solution to the above equation is
\be \label{e44}
\Omega = {a\over m (\cosh\alpha + \cosh\beta) (m + \sqrt{m^2 - a^2})}\, .
\ee
{}From eqs.\refb{e41}, \refb{e42} we see that
in the extremal limit $a\to m$,
\be \label{e45}
A\to 4\pi m^2(\cosh\alpha + \cosh\beta) =
8\pi |J|\, , \qquad \kappa\to 0\, .
\ee

For non-rotating black holes ($a=0$) special care is needed to study the
extremal limit. {}From eqs.\refb{e34}, \refb{e27} we see that in this case,
\be \label{e46}
ds_E^2 = -\Delta^{-{1\over 2}} (\rho^2 - 2m\rho) dt^2 + \Delta^{1\over 2}
(\rho^2 - 2m\rho)^{-1} d\rho^2 + \Delta^{1\over 2} (d\theta^2 +
\sin^2\theta d\phi^2)\, ,
\ee
where,
\be \label{e47}
\Delta = \rho^2 \{ \rho^2 +2m\rho(\cosh\alpha\cosh\beta -1) + m^2
(\cosh\alpha - \cosh\beta)^2\}\, .
\ee
The metric describes a black hole with horizon at $\rho=2m$ and singularity
at $\rho =0$. The extremal limit corresponds to the case when the horizon
approaches the singularity, {\it i.e.} $m\to 0$ keeping the physical mass
$M$ defined in eq.\refb{e35} fixed. We shall now study this limit in three
separate cases.

\noindent Case I: $\alpha>\beta$

In this case we consider the limit
\be \label{e48}
m\to 0, \qquad \alpha\to \infty, \qquad m\cosh\alpha\equiv m_0 \, \,
\hbox{finite}, \qquad \beta \, \, \hbox{finite}\, .
\ee
Then eqs.\refb{e35} and \refb{e37} take the form:
\be \label{e49}
M={m_0\over 2} \cosh\beta, \qquad \vec Q_L = {m_0\over \sqrt 2}
\cosh\beta \, \vec n, \qquad \vec Q_R = {m_0\over \sqrt 2}
\sinh\beta \,  \vec p\, .
\ee
$\vec Q_L$ and $\vec Q_R$ are 22 and 6 dimensional vectors respectively, and
have been defined below eq.\refb{e51}.
{}From this we get the following relations between
$M$, $\vec Q_L$ and $\vec Q_R$:
\be \label{e52}
M^2= {1\over 2} \vec Q_L^2\, , \qquad \vec Q_R = \sqrt 2 M \tanh\beta \,
\vec p\, .
\ee
Thus here $\vec Q_R^2<\vec Q_L^2$. {}From eqs.\refb{e41} and \refb{e42}
we see that in this limit,
\be \label{eext1}
A\to 0, \qquad \kappa \to {1\over 4M} \cosh\beta\, .
\ee

\noindent Case II: $\alpha<\beta$

In this case we consider the limit
\be \label{e53}
m\to 0, \qquad \beta \to \infty, \qquad m\cosh\beta \equiv m_0 \, \,
\hbox{finite}, \qquad \alpha \, \, \hbox{finite}\, .
\ee
In this limit, eqs.\refb{e35} and \refb{e37} take the form:
\be \label{e54}
M={m_0\over 2} \cosh\alpha\, , \qquad \vec Q_L = {m_0\over \sqrt 2}
\sinh\alpha \, \vec n\, , \qquad \vec Q_R = {m_0\over \sqrt 2}
\cosh\alpha \, \vec p\, .
\ee
Thus now,
\be \label{e55}
M^2 = {1\over 2} \vec Q_R^2\, , \qquad \vec Q_L = \sqrt 2 M\tanh\alpha
\, \vec n\, .
\ee
In this limit $\vec Q_L^2 < \vec Q_R^2$.
Also,
\be \label{eext2}
A\to 0, \qquad \kappa\to {1\over 4M} \cosh\alpha\, .
\ee

\noindent Case III: $\alpha=\beta$.

In this case we consider the limit:
\be \label{e56}
m\to 0, \qquad \alpha=\beta\to \infty, \qquad m\cosh^2\alpha\equiv m_0 \,
\, \hbox{finite}\, .
\ee
In this limit,
\be \label{e57}
M={m_0\over 2}, \qquad \vec Q_L = {m_0\over \sqrt 2} \, \vec n, \qquad
\vec Q_R = {m_0\over \sqrt 2} \, \vec p\, .
\ee
Thus,
\be \label{e58}
M^2 = {1\over 2} \vec Q_L^2 = {1\over 2} \vec Q_R^2\, .
\ee
Also in this limit,
\be \label{eext3}
A\to 0, \qquad \kappa\to \infty\, .
\ee

In the normalization convention that we have been using, the Bogomol'nyi
bound that follows from the space-time supersymmetry of the theory,
is given by $M^2\ge (\vec Q_R^2/2)$\cite{BOGOM,BOGOMS}. Thus we see that the
extremal black holes in cases II and III saturate the Bogomol'nyi bound,
and hence give rise to supersymmetric solutions,
whereas those in case I do not. Also for non-zero $J$,
eqs.\refb{e35}-\refb{e37} shows that $M^2$ is always larger than
$(\vec Q_R^2/2)$, even in the extremal limit $a\to m$. This shows that
the extremal black holes carrying non-zero angular momentum do not
correspond to supersymmetric solutions. (Similar observations have been
made in ref.\cite{KALLTWO}.)
For special cases $\alpha=\beta$ and $\alpha=0$, the non-rotating
solutions constructed here reproduce the solutions
discussed in ref.\cite{DUFF}.

\sectiono{Rotating Dyonic Black Holes} \label{sthree}

In this section we shall discuss construction of black holes carrying
both electric and magnetic charges. One class of dyons that can be
constructed trivially are the ones for which the magnetic charge vector
$\vec Q_{mag}$ is parallel to $L\vec Q_{el}$, where $\vec Q_{el}$ is
the electric charge vector. This is done with the help of SL(2,R)
transformations\cite{DEROO,STW,SLTZ,SCHSE}.
To see  how this works, we first note
that the equation of motion for the $B_{\mu\nu}$ field, derived from
the action \refb{e1}, allows us to introduce a field $\Psi$, related to
$H_{\mu\nu\rho}$ through the relation:
\be \label{ea2}
g^{\mu\mu'} g^{\nu\nu'} g^{\rho\rho'} H_{\mu'\nu'\rho'}
= - (-\det g)^{-{1\over 2}} e^{2\Phi} \epsilon^{\mu\nu\rho\sigma}
\p_\sigma \Psi\, , \qquad g_{\mu\nu} \equiv e^{-\Phi} G_{\mu\nu}\, .
\ee
Let us now define the complex scalar field:
\be \label{ea1}
\lambda = \Psi + i e^{-\Phi} \equiv \lambda_1 + i\lambda_2\, .
\ee
Then the equations of motion and the Bianchi identities of the theory can
be shown to be invariant under the transformations:
\ben \label{ea4}
\lambda &\to& {a\lambda + b\over c\lambda+d}, \qquad g_{\mu\nu}
\to g_{\mu\nu}, \qquad M\to M\, , \nonumber \\
F^\ab_{\mu\nu} &\to& (c \lambda_1+d) F^\ab_{\mu\nu} + c\lambda_2
(ML)_{ab}\wt F^\bb_{\mu\nu}\, ,
\qquad \qquad ad - bc=1, \quad a,b,c,d\in {\bf R}\,  ,
\een
where,
\be \label{ea4a}
\wt F^\ab_{\mu'\nu'} = {1\over 2} (-\det g)^{-{1\over 2}}
g_{\mu\mu'} g_{\nu\nu'} \epsilon^{\mu\nu
\rho\sigma} F^\ab_{\rho\sigma}\, .
\ee
Thus given a solution of the classical equations of motion, we can
generate a new solution by performing the above SL(2,R) transformation
on the original solution.

In accordance with our earlier spirit we shall look for solutions with
fixed asymptotic values of various fields. Let us choose the asymptotic
values of $\Phi$ and $\Psi$ to be zero. A solution with non-zero
asymptotic values of $\Phi$ and $\Psi$ can be brought to this form
by using the freedom of shifting $\Phi$ and $\Psi$ by
constants keeping $G_{\mu\nu}$, $A^\ab_\mu$ and $M$ fixed. Thus we do
not suffer from any loss of generality by restricting our solutions this
way. This forces us to consider only those SL(2,R)
transformations which leave the point $\Phi=\Psi=0$ fixed. These
transformations belong to an SO(2) subgroup of SL(2,R), and are
represented by the matrix
\be \label{ea4b}
\pmatrix{a&b\cr c&d\cr} = \pmatrix{\cos\gamma & \sin\gamma \cr -
\sin\gamma & \cos\gamma\cr}\, .
\ee
We can now apply this transformation to the electrically charged rotating
black hole solution given in eqs.\refb{e26}-\refb{e34}. Since $g_{\mu\nu}$
does not transform under the SL(2,R) transformation, the geometry remains
identical. The matrix valued scalar field $M$ also remains the same as
given in eqs.\refb{e32}, \refb{e33}.
The fields $\Phi$, $A_\mu^\ab$ and $B_{\mu\nu}$ (or
equivalently $\Psi$) change. We shall not write down the transformed
solutions explicitly, but only give the asymptotic forms of the gauge
field strengths which allow us to extract the electric and the magnetic
charges associated with this solution. These are given by,
\ben \label{ea5}
F^\ab_{\rho t} &\simeq& {n^\ab\over \sqrt 2} {m\over \rho^2} \sinh\alpha
\cosh\beta \cos\gamma \quad \hbox{for} \, \, 1\le a\le 22\, , \nonumber \\
&\simeq& {p^{(a-22)}\over \sqrt 2} {m\over \rho^2} \sinh\beta \cosh\alpha
\cos\gamma \quad \hbox{for} \, \, 23 \le a \le 28\, ,
\een
\ben \label{ea6}
\wt F^\ab_{\rho t} &\simeq& -{n^\ab\over \sqrt 2} {m\over \rho^2} \sinh\alpha
\cosh\beta \sin\gamma \quad \hbox{for}\, \, 1\le a\le 22\, , \nonumber \\
&\simeq& {p^{(a-22)}\over \sqrt 2} {m\over \rho^2} \sinh\beta \cosh\alpha
\sin\gamma \quad \hbox{for} \, \, 23 \le a \le 28\, .
\een
This gives,
\ben \label{ea7}
Q^\ab_{el} &=& {n^\ab\over \sqrt 2} \, m \sinh\alpha
\cosh\beta \cos\gamma \quad \hbox{for} \, \,  1\le a\le 22\, , \nonumber \\
&=& {p^{(a-22)}\over \sqrt 2} \, m \sinh\beta \cosh\alpha
\cos\gamma \quad \hbox{for} \, \,  23 \le a \le 28\, ,
\een
\ben \label{ea8}
Q^\ab_{mag} &=& -{n^\ab\over \sqrt 2} \, m \sinh\alpha
\cosh\beta \sin\gamma \quad \hbox{for} \, \,  1\le a\le 22\, , \nonumber \\
&=& {p^{(a-22)}\over \sqrt 2} \, m \sinh\beta \cosh\alpha
\sin\gamma \quad \hbox{for} \, \,  23 \le a \le 28\, .
\een
Although the above solution represents a rotating dyonic black hole
solution,
it does not correspond to a black hole with a general electric and
magnetic charge vector. In particular, the electric and the magnetic
charge vectors are related as
\be \label{ea16}
Q^\ab_{mag} = \tan\gamma L_{ab} Q^\bb_{el}\, .
\ee

We shall now discuss the construction of a rotating black hole solution
carrying independent electric and magnetic charge vectors. The basic
strategy that we employ is to make successive use of SL(2,R) and
O(7,23) transformations. Since these two sets of transformations do not
commute, we would expect to produce solutions carrying charges that are
more general than the ones given in eqs.\refb{ea7}, \refb{ea8} by using
this procedure. However, as pointed out in ref.\cite{HASSAN}, there is
a potential problem with this approach. The magnetically
charged solution is characterized by an $A^\ab_\mu$ which is not
globally defined, but need to be defined using two different coordinate
patches. Now, if we perform a general O(7,23) transformation on this
solution, the component $G_{ti}$ of the metric mixes with $A^\ab_i$.
Thus in the resulting field configuration, the metric will not be globally
defined, but need to be defined in two separate coordinate patches, one
around the positive $z$-axis, and the other around the negative $z$-axis.
On the overlap of the two coordinate patches, the two
solutions are related by a coordinate transformation of the form:
\be \label{etphi}
t\to t + c\phi\, ,
\ee
for some constant $c$. This, however is not a globally defined coordinate
transformation, since it is not invariant under $\phi\to \phi+2\pi$.
Hence the solution constructed this way would not be an acceptable
field configuration. Alternatively, one could use only one coordinate
patch, but then the resulting metric will have a singularity either on
the positive or on the negative $z$-axis.

This shows that we have to be careful about applying SL(2,R) and
O(7,23) transformations on a solution successively, but it does not
rule out the possibility that suitable combinations of these
transformations can be found which give rise to a non-singular metric.
We shall now show how this can be done in a systematic manner to produce
a 58 parameter black hole solution carrying arbitrary mass, angular
momentum, electric charge and magnetic charge. The first point to note
is that since for time independent solutions the equations of motion of this
theory are the same as those of the ten dimensional
heterotic string theory for field configurations independent of seven
of the ten dimensions (the time and the six internal coordinates), these
equations of motion are expected to have an O(8,24)
symmetry\cite{MARSCH,THREED}.\footnote{Actually, since the Kerr solution
is independent of two coordinates, $t$ and $\phi$, there is in fact an
infinite parameter Geroch group of transformations that can be used to
generate new solutions\cite{BAKAS}. However, most of these solutions
are singular, and so we restrict ourselves to the O(8,24) group of
transformations.} To see how this O(8,24) symmetry manifests
itself in the present case, let us first note that the equations of
motion of the $\wb A^\bab_i$ fields, derived from the action \refb{e13}
gives,
\be \label{e61}
\p_i \Big (\sqrt{\det\wb G} e^{-\wb \Phi} (\wb M \wb L)_{\wb a \wb b}
\wb G^{ii'} \wb G^{jj'} \wb F^\bbb_{i'j'}\Big)=0\, ,
\qquad 1\le\wb a\le 30\, .
\ee
This allows us to introduce a set of fields $\psi^{\wb a}$ through
the equations
\be \label{e62}
\sqrt{\det\wb G} e^{-\wb \Phi} (\wb M \wb L)_{\wb a \wb b}
\wb G^{ii'} \wb G^{jj'} \wb F^\bbb_{i'j'} = {1\over 2} \epsilon^{ijk}
\p_k \psi^{\wb a}\, .
\ee
The bianchi identity for the gauge fields, $\epsilon^{ijk}\p_i
\wb F^\bab_{jk} = 0$ gives,
\be \label{e63}
\wb D^i (e^{\wb \Phi} (\wb M \wb L)_{\wb a \wb b} \p_i \psi^{\wb b})=0
\, ,
\ee
where $\wb D_i$ denotes the covariant derivative with respect to the metric
$\wb G_{ij}$. Let us now regard $\psi$ as a 30 dimensional column vector
and define,
\be \label{e64}
\MM = \pmatrix{\wb M - e^{2\wb \Phi} \psi \psi^T & e^{2\wb \Phi} \psi
& \wb M \wb L \psi -{1\over 2} e^{2\wb \Phi} \psi (\psi^T \wb L \psi)
\cr e^{2\wb\Phi} \psi^T & - e^{2\wb\Phi} & {1\over 2} e^{2\wb \Phi}
\psi^T \wb L \psi \cr \psi^T \wb L \wb M - {1\over 2} e^{2\wb \Phi}
\psi^T(\psi^T \wb L \psi) & {1\over 2} e^{2\wb\Phi} \psi^T \wb L \psi
& - e^{-2\wb\Phi} +\psi^T \wb L \wb M \wb L \psi -{1\over 4}
e^{2\wb\Phi} (\psi^T \wb L \psi)^2 \cr }\, ,
\ee
\be \label{e65}
\LL = \pmatrix{ \wb L & 0 & 0\cr 0 & 0 & 1\cr 0 & 1 & 0\cr}\, ,
\ee
and,
\be \label{e63b}
\wb g_{ij} = e^{-2\wb \Phi} \wb G_{ij}\, .
\ee
If we restrict ourselves to field configurations for which $\wb H_{ijk}=0$,
then eq.\refb{e63} and the other equations of motion derived from the action
\refb{e13} can be shown to be identical to the equations of motion
derived from the action:
\be \label{e63a}
\wt S = \wt C \int d^3 x \sqrt{\det \wb g} \, [ R_{\wb g} +{1\over 8}
\wb g^{ij} Tr(\p_i \MM \LL \p_j \MM \LL) ]\, .
\ee
$\wt S$ is manifestly invariant under the O(8,24) transformation:
\be \label{e66}
\MM\to \wt \Omega \MM \wt\Omega^T\, , \qquad \wb g_{ij} \to \wb g_{ij}\, ,
\ee
where $\wt \Omega$ is a 32$\times$32 matrix satisfying
\be \label{e67}
\wt\Omega \LL \wt \Omega^T = \LL\, .
\ee

We can use this O(8,24) transformation to generate new solutions of the
equations of motion from a given time independent solution.
As before, we shall consider only a subgroup of this O(8,24) group of
transformations
which keeps the asymptotic field configuration fixed. Together with
the asymptotic conditions on $G_{\mu\nu}$, $\Phi$ and $M$ that we have
already imposed, if we further restrict the $\psi^{\wb a}$'s to vanish
asymptotically\footnote{Note that adding a constant to $\psi^{\wb a}$
does not
change the equations of motion, and hence any non-zero constant asymptotic
value of $\psi^{\wb a}$ can be removed by simply subtracting that constant
from $\psi^{\wb a}$ in the solution.} then the asymptotic value of
$\MM$ is given by,
\be \label{e68}
\MM_{as} = \pmatrix{I_{28} & \cr & -I_4 \cr}\, .
\ee
To see what subgroup of O(8,24) transformations preserve $\MM_{as}$, let
us work in a representation in which $\LL$ is diagonal. The orthogonal
matrix $\UU$ which diagonalizes $\LL$ is,
\be \label{e70}
\UU = \pmatrix{I_{28} & 0 & 0 & 0 & 0 \cr 0 & {1\over \sqrt 2}
& {1\over \sqrt 2} & 0 & 0 \cr 0 &
{1\over \sqrt 2} & -{1\over \sqrt 2}
& 0 & 0 \cr 0 & 0 & 0 & {1\over \sqrt 2} & {1\over \sqrt 2} \cr
0 & 0 & 0 &
{1\over \sqrt 2} & -{1\over \sqrt 2} \cr }\, .
\ee
We have,
\be \label{e69}
\UU \LL \UU^T = \LL_d \, ,
\ee
where,
\be \label{e71}
\LL_d = \pmatrix{ -I_{22}&&&&& \cr  & I_6 &&&& \cr && 1 &&& \cr &&& -1 && \cr
&&&& 1 & \cr
&&&&& -1 \cr }\, .
\ee
{}From eqs.\refb{e68} and \refb{e70} we get,
\be \label{e72}
\UU \MM_{as} \UU^T = \pmatrix{I_{28} & \cr & -I_4 \cr}\, .
\ee
Thus the matrices $\UU\wt \Omega \UU^T$, which preserve both
$\LL_d$ and $\UU \MM_{as} \UU^T$ belong to an
O(22,2)$\times$O(6,2) transformations. The O(22,2) transformation acts
on the 1st - 22nd, 29th and 31st row, whereas the O(6,2) transformation
acts on the 23rd - 28th, 30th and 32nd row.

We shall now study the effect of these transformations on the original
Kerr solution \refb{e22}. The
metric $\wb g_{ij}$ and matrix $\MM$ for this solution can be
easily computed, and the result is,
\be \label{e74b}
\wb g_{ij} dx^i dx^j = (\rho^2 + a^2\cos^2\theta -2m\rho) \Big[
{1 \over \rho^2 + a^2 -2m\rho} d\rho^2 + d\theta^2 + {\rho^2 + a^2 -2m\rho
\over \rho^2 + a^2 \cos^2\theta - 2m\rho} \sin^2\theta d\phi^2\Big]\, ,
\ee
\be \label{e74}
\MM = \pmatrix{I_{28} & 0 & 0 & 0 & 0 \cr 0 & -f^{-1} & 0 & 0 & -g \cr
0 & 0 & -f - f g^2 & g & 0 \cr 0 & 0 & g & - f^{-1} & 0 \cr
0 & -g & 0 & 0 & -f - f g^2 \cr } \equiv \MM^\zero \, ,
\ee
where,
\ben \label{e74a}
f & = & (\rho^2 + a^2 \cos^2 \theta - 2m\rho)/(\rho^2 + a^2 \cos^2\theta)
\nonumber \\
g & = & 2ma\cos\theta / (\rho^2 + a^2 \cos^2\theta - 2m\rho)\, .
\een
It is clear that the matrix $\UU\MM^\zero\UU^T$ is left invariant under the
O(22)$\times$O(6) subgroup of the O(22,2)$\times$O(6,2) group, which
act on the first 22 and the 23rd-28th indices of the matrix respectively.
What is perhaps not so obvious is that this matrix is also left invariant
under an SO(2) subgroup of
O(22,2)$\times$O(6,2), represented by the matrix
\be \label{enew1}
\UU \wt \Omega \UU^T = \pmatrix{I_{28} &&&& \cr & \cos\alpha & 0 & \sin\alpha
& 0 \cr
& 0 & \cos\alpha & 0 & \sin
\alpha \cr & -\sin\alpha & 0 & \cos\alpha & 0 \cr & 0 & -\sin\alpha & 0
& \cos\alpha\cr}\, .
\ee
This can be verified by explicit computation. Thus the transformations
which generate inequivalent solutions are parametrized by the coset:
\be \label{e76}
\big(O(22,2) \times O(6,2)\big) / \big(O(22)\times O(6)
\times SO(2)\big)\, .
\ee
This coset is parametrized by 57 parameters. This, together with the
original parameters $m$ and $a$, gives a 59 parameter solution. But we
now recall that not all of these solutions are non-singular, due to
the possible singularity in the $G_{t\phi}$ component of the metric
discussed previously. The presence of such a singularity is signalled
by an asymptotic value of $\p_\theta G_{t\phi}$ proportional to $\sin
\theta$, in the same way that the presence of magnetic charge associated
with the gauge field $A^\ab_\mu$ is signalled by the presence of an
asymptotic value of $F^\ab_{\theta\phi}$ proportional to $\sin\theta$.
{}From eqs.\refb{e7} and \refb{e62} we see that such an asymptotic form of
$G_{t\phi}$ would induce an asymptotic $\psi^{30}$ proportional to
$1/\rho$. Thus in order to get a non-singular metric, we must demand
that the coefficient of $1/\rho$ in the
asymptotic expression of $\psi^{30}$ must vanish. (This is equivalent
to demanding that the coefficient $c$ appearing in eq.\refb{etphi}
vanishes.) This gives one constraint among the 59
parameters, thereby reducing the number of independent parameters to
58. Using eqs.\refb{e64}, \refb{e66}, \refb{e74} and \refb{e74a} we can
derive an explicit form of this constraint on the O(22,2)$\times$O(6,2)
matrix $\wt \Omega$:
\be \label{econs1}
\wt \Omega_{30,29} \wt\Omega_{31,29} + \wt \Omega_{30,31} \wt\Omega_{31,31}
- \wt\Omega_{30,30} \wt\Omega_{31,30} - \wt\Omega_{30,32} \wt\Omega_{31,32}
=0\, .
\ee

Note that 58 is precisely the expected number of parameters required to
label the most general static black hole solution consistent with the
no hair theorem, since such a black hole will be characterized by
mass, angular momentum, 28 electric
charges and 28 magnetic charges. Thus
although we have not explicitly shown that the solutions constructed
this way are free from naked singularities, there is good
reason to believe that this is indeed the case.

We shall end this section by giving an interpretation of the 59th
parameter that takes us out of the class of non-singular solutions.
A typical representative O(8,24) transformation which does not satisfy
\refb{econs1}, and hence, acting on a
Kerr solution, produces a singular solution,  is given by,
\be \label{eb1}
\UU\wt \Omega \UU^T = \pmatrix{I_{28} &&&& \cr
& \cos\alpha & 0 & \sin\alpha & 0 \cr
& 0 & 1 & 0 & 0  \cr
& -\sin\alpha & 0 & \cos\alpha & 0 \cr
& 0 & 0 & 0 & 1  \cr}\, .
\ee
For simplicity, let
us restrict ourselves to studying the effect of this transformation on
the Schwarzschild solution ($a=0$).
In this case, the transformed solution is given by,
\ben \label{eb2}
G_{\mu\nu} dx^\mu dx^\nu &=& -{\rho(\rho-2m)\over \wt
\Delta} (dt + m \sin\alpha
\cos\theta d\phi)^2 + {\wt\Delta \over \rho(\rho-2m)} d\rho^2
+ \wt\Delta (d\theta^2 + \sin^2\theta d\phi^2)\, ,
\nonumber \\
\Phi &=& 0\, \qquad A_\mu^\ab =0, \qquad M=I_{28}, \qquad B_{\mu\nu}=0 \, ,
\een
where,
\be \label{eb3}
\wt\Delta = \rho^2 - 4m\rho \sin^2{\alpha\over 2} + 4m^2 \sin^2{\alpha\over 2}
\, .
\ee
This can be easily recognized as the Taub-NUT solution\cite{TAUB}.
The action of the other elements of the O(8,24) group then produces the
Taub-NUT dyon solutions discussed in refs.\cite{KALLTWO,MYER}.
For non-zero $a$, the solutions generated this way correspond to
rotating Taub-NUT dyon solutions given in ref.\cite{KECH}.

Thus we see that the rotating dyonic Taub-NUT solutions
can also be generated via O(8,24) transformation of the
original Kerr solution. This has already been observed
before\cite{KECH,BAKAS2,MYERS2}, and can be traced to the fact that
the transformation on the fields represented by the
O(8,24) matrix \refb{eb1} is precisely the Ehlers-Geroch
transformation\cite{GEROCH}
that takes us from the Schwarzschild solution to the Taub-NUT family of
solutions.
To this effect, we note that the scalar field $\psi^{30}$ defined through
eqs.\refb{e62}, \refb{e7} is simply the generalization of the NUT
potential\cite{GIBHAW}, and requiring it to fall off faster than $1/\rho$
asymptotically corresponds to setting the NUT charge to zero.

\sectiono{Conclusion} \label{sfour}

In this paper we have explicitly constructed the general electrically
charged rotating black hole solution in heterotic string theory compactified
on a six dimensional torus. We have also given
an algorithm to construct the general rotating black hole solution
carrying both, electric and magnetic charges. This is the most general
black hole solution in this theory consistent with no hair theorem.

We hope that these results can be used to study the relationship between
black holes and elementary string states in the spirit of
refs.\cite{SUSSKIND,RUSSO}. To this end, we shall conclude this paper by
pointing out some amusing coincidences between the
properties of black holes
and those of elementary strings in the extremal limit. First we compare
the gyromagnetic ratios of
the solutions found in this paper
and those of elementary string states in the limit
$\beta\to \infty$ with the physical mass $M$, the physical angular
momentum $J$, and the parameter $\alpha$ fixed. Here $\alpha$ and $\beta$
are the parameters appearing in the black hole solution given in section
\ref{stwo}.
(Note that in this limit the solution develops naked singularity, but
we shall ignore this problem. We can take the limit $J\to 0$ at the end of
the calculation, so that the solution approaches an extremal black hole.)
{}From eq.\refb{emu2} we see that in this limit,
\be \label{ecomp1}
g_L\to 0, \qquad g_R\to 2\, .
\ee
On the other hand, these gyromagnetic ratios for elementary string states
can be computed using a slight generalization of the work of
ref.\cite{RUSSO}. The answer is,
\be \label{ecomp2}
g_L = 2 \, {S_R\over S_R + S_L}\, , \qquad g_R = 2 \, {S_L\over S_R
+ S_L}\, ,
\ee
where $S_L$ and $S_R$ denote the contribution to the $z$ component of the
angular momentum from the left and the right moving oscillators
respectively. From eqs.\refb{e35} - \refb{e37} we see that in the limit
$\beta\to \infty$, with $M$, $J$ and $\alpha$ fixed, $M^2-\vec Q_R^2/2
\to 0$. Thus these states saturate the Bogomol'nyi bound, which, in
turn, implies that the right hand part of this state must be the lowest
state in the Neveu-Schwarz (or Ramond) sector consistent with GSO
projection\cite{BOGOMS}. Thus for these states $|S_R|\le\hbar$, and most
of the contribution to $J$ comes from $S_L$.
This shows that in this limit $|S_R|<<|S_L|$, and $g_L$ and $g_R$
given in eq.\refb{ecomp2} agree with those given in eq.\refb{ecomp1}.

Next we shall compare the area of the stretched horizon and the logarithm
of the density of single string states in the extremal limit. We shall
restrict ourselves to the non-rotating solution since in this case the
extremal limit \refb{e53} corresponds to saturation of the Bogomol'nyi
bound, and various non-renormalization theorems are expected to be
valid. As we can see from eq.\refb{eext2}, the area of the event horizon
vanishes in the limit \refb{e53}. However, if
$A_s$ denotes the area of the surface $\rho=\rho_H^+ + \eta$, where $\eta$
is a fixed length of the order of the Planck length, then using
eq.\refb{e46} we find that it is non-zero and has the value,
\be \label{efinal1}
A_s = 8\pi M\eta/\cosh\alpha = 8\pi \eta \sqrt{M^2 - {\vec Q_L^2\over 2}}
\, .
\ee
This surface is similar to the stretched horizon discussed in
refs.\cite{SUSSKIND,RUSSO}. On the other hand, the density of states of
elementary string excitations
in the extremal limit $M^2\to \vec Q_R^2/2$ can be
easily calculated (see, for example, ref.\cite{RUSSO})
and is given by,
\be \label{efinal2}
d_S \sim \exp(4\pi \sqrt{n})\, ,
\ee
where $n$ denotes the total contribution to mass$^2$ of the state from the
left moving oscillators. (Note that in this case there is no appreciable
contribution to the density of states from the right moving oscillators.)
This, in turn, is proportional to
$M^2 - (\vec Q_L^2/2)$. Thus we see that,
\be \label{efinal3}
A_s \propto \ln d_S\, .
\ee
This establishes the relationship between the density of string states and
the area of the stretched horizon in the extremal limit.

\end{document}